\documentclass[10pt,conference]{IEEEtran}
\IEEEoverridecommandlockouts
\usepackage{cite}
\usepackage{amsmath, amssymb, amsfonts}
\usepackage{mathtools}

\usepackage{cuted}
\usepackage[utf8]{inputenc}
\usepackage[T1]{fontenc}
\usepackage{graphicx}
\usepackage{adjustbox}

\usepackage{textcomp}
\usepackage{xcolor}
\usepackage{cuted}
\usepackage{amsmath, amssymb, amsthm,algorithm}
\usepackage{latexsym}
\usepackage{graphics,epsfig}
\usepackage{amsfonts}
\usepackage{booktabs}
\usepackage{color}
\usepackage{multirow}
\usepackage[justification=centering]{caption}
\usepackage{dsfont}
\usepackage{graphicx}
\usepackage{adjustbox}
\usepackage{kantlipsum}
\usepackage[caption=false, font=footnotesize, labelfont=sf, textfont=sf]{subfig}
\usepackage{cases}
\usepackage{url}
\usepackage{makecell}
\usepackage{soul}

\usepackage{array}
\usepackage{algpseudocode}
\usepackage{mathtools,lipsum,cuted}

\usepackage[justification=centering]{caption}

\usepackage[caption=false,font=footnotesize]{subfig}
\usepackage[T1]{fontenc}
\usepackage{mwe}
\usepackage{subfig}
\usepackage{pdflscape}

\newlength{\tempheight}
\newlength{\tempwidth}

\newcommand{\rowname}[1]
{\rotatebox{90}{\makebox[\tempheight][c]{#1}}}

\newcommand{\columnname}[1]
{\makebox[\tempwidth][c]{#1}}

\makeatletter
\def\BState{\State\hskip-\ALG@thistlm}
\makeatother

\newcounter{example}[section]

\usepackage[english]{babel}
\usepackage{verbatim}

\usepackage[english]{babel}

\theoremstyle{plain}

\theoremstyle{remark}

\def\BibTeX{{\rm B\kern-.05em{\sc i\kern-.025em b}\kern-.08em
    T\kern-.1667em\lower.7ex\hbox{E}\kern-.125emX}}
\begin{document}

\voffset=0.05in
\textheight=9.28in

\title{Optimizing Split Learning Latency in TinyML-Based IoT Systems}

\author{\IEEEauthorblockN{Zied Jenhani, Mounir Bensalem, Jasenka Dizdarevi\'c   and Admela Jukan}
\IEEEauthorblockA{Technische Universit\"at Braunschweig, Germany;
\{zied.jenhani, mounir.bensalem,  j.dizdarevic, a.jukan\}@tu-bs.de}
}

\maketitle

\begin{abstract}
Split learning (SL) addresses the limitation of running deep learning inference directly on low-power edge/IoT nodes, in which it executes part of the inference process on the sensor and offloading the remainder to a companion device. Despite its promise, the inference latency of SL on constrained hardware under realistic low-power wireless protocols 
remains unexplored. This paper presents the first experimental latency benchmark of TinyML-based SL on ESP32-S3 boards, comparing four wireless communication protocol solutions (UDP, TCP, ESP-NOW, BLE).  We also analyze the impact of the choice of different split points across different models (MobileNet-V2 and ResNet50) in terms of communication and computation overhead as a way to minimize the end-to-end inference latency. We propose a Beam Search–based algorithm for split point optimization that minimizes end-to-end latency, and compare it with other methods, including Greedy Search, First-Fit, Random-Fit, and Brute Force. ESP-NOW achieves the best RTT (3.6 s) and serves as the base protocol for the algorithm, which delivers near-optimal latency with processing time of 0.1 s for 5 devices.

\end{abstract}


\section{Introduction}\label{sec:intro}

The next innovation frontier in the rapid evolution of AI is in its ability to run deep learning inference on ultra-low power resource-constrained edge/IoT devices \cite{10929047}. Significant advancements towards this goal are emerging, with the paradigm shift towards on-device computing, enabled by optimizations of deep learning (DL) models \cite{10.1145/3657282}. TinyML models present one of the most promising solutions, as they provide lightweight inference for resource-constrained hardware \cite{10.1145/3661820}. Furthermore, Split Learning (SL) can be used to overcome processing power and memory size limitations by 
partitioning a model across devices,  early layers can run on the sensor, whereas the remaining layers can execute on a 
companion microcontroller (MCU) or edge server \cite{li2024adaptive}. On-device inference preserves privacy and assures bandwidth benefits, since only intermediate activations are exchanged. 

\par Although SL has been analyzed on smartphones and single board computers \cite{matsubara2022split}, empirical evidence on low-power microcontrollers is lacking, particularly regarding the role of different wireless protocols in end-to-end latency. The performance measurements and evaluations are missing in the context of split latency incurred, in terms of processing time, including network and communication setup, ML inference, and the transmission of intermediate activations between IoT devices. In addition, RTT and delay associated with the transfer of intermediate activations and prediction data delay between devices is still unknown, particularly in terms of optimal split point choice, which requires new studies \cite{trung2025latency}. Finally, the impact of over-the-air communication modes on system performance, such as when using different protocols is an open issue. Communication networks based on WiFi, ESP-NOW, and BLE are all expected to yield different performance.

In this work, we examine the potential of implementing the SL approach in resource-constrained environment, by experimentally benchmarking SL TinyML inference latency on low-cost open-source ESP32-based IoT devices. For the implementation of TinyML framework we use  TensorFlow Lite (TFLite), as it comes with sufficient developer support. The model is prepared (partitioned and quantized) in an edge server for which we use a Desktop PC, from which the firmware is then remotely deployed to the IoT devices using Over-the-Air (OTA) firmware updates without requiring the physical access. Additionally, we introduce a  novel approach to optimize SL split points, based on the Beam Search algorithm and study the same experimentally. 
The experimental results in the manually chosen split point at \texttt{block\_16\_project\_BN} layer demonstrate the best results, regarding end-to-end inference latency for the use of ESP-NOW protocol (with the RTT of 3.6 s) compared to BLE, TCP/IP and UDP/IP. Consequently, we used ESP-NOW as a base solution for the Beam Search-based algorithm. The results of the algorithm demonstrated the significantly lower processing times (around 0.1 s for 5 devices) while maintaining competitive latency performance.\


\par The rest of this paper is organized as follows:
Section~\ref{sec:related} surveys related work. Section~\ref{sec:system}
describes the system architecture. Section~\ref{sec:solution} describes the problem formulation and optimization framework. Section~\ref{sec:results} presents and discusses the measurements. Section~\ref{sec:conclusion} concludes the paper.

\section{Related Work}\label{sec:related}

TinyML enables AI inference on resource-constrained IoT devices \cite{10929047}, with ESP32 boards being a common deployment target due to their open-source design and cost effectiveness
\cite{10.1145/3661820}. Combined with Split Learning (SL), TinyML can further reduce latency and power consumption, making it a key enabler for constrained AI applications
\cite{10454775,hafi2024split}.

 

Existing work centers around improving model accuracy and reducing communication overhead for traditional servers or edge devices \cite{ko2023dynamic}. One of the critical challenges in SL is the manual or heuristic-based selection of split points.  A recent study proposed a genetic algorithm (GA)-based solution to optimize split point selection \cite{trung2025latency}, assuming that a model is split between two devices only, and that the transmission delay depends on the size of the data and the data rate. Our work introduces for the first time Beam Search method for automatic split point selection in SL, with the objective to minimizing end-to-end inference latency, i.e., accounting for potential effects of communication overhead due to protocol settings and packet loss, which related work does not consider. We adopt Beam Search algorithm due to its fast convergence as compared to GA, which typically needs a large number of population 
and generation. 
Beam Search can provide a balance between exhaustive search accuracy and greedy heuristic speed by exploring only a limited number of promising candidate partitions at each step \cite{huber2021learning}.

Considering the adoption of communication protocols for SL, the work in \cite{10454775}  have evaluated a single WiFi configuration. We provide a latency analysis of four wireless networking protocols (UDP, TCP, ESP-NOW, and BLE) to inform the design of future TinyML + SL deployments. The source code used in our experiments is publicly available, providing a reproducible baseline for researchers and developers alike.


\section{System Architecture}\label{sec:system}

\begin{figure}[ht]
\centering
\includegraphics[width=\linewidth]{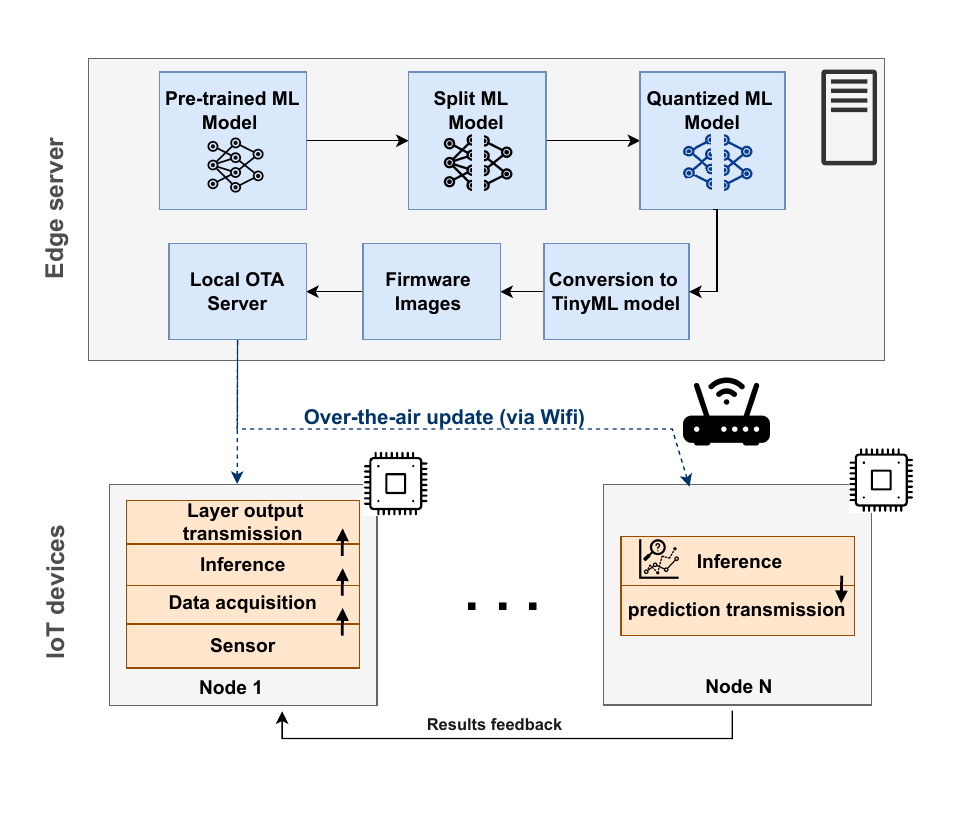}
\caption{SL-TinyML framework architecture communication process}
\label{fig:arch1}
\end{figure}

Fig.~\ref{fig:arch1} shows the system architecture for SL-TinyML integration framework solution for the better utilization of computing resources distributed across edge/IoT nodes. In the proposed framework, computation and processing tasks are separated in two contextual layers, IoT and the edge, with the edge server side used for model training, model partitioning, post-training quantization, firmware generation and  OTA updating of firmware images for the IoT devices, across which the inference is distributed.  

As depicted in Fig.~\ref{fig:arch1}, the initial step is the deployment of a pre-trained, quantized ML model on an edge server. The model is then divided into multiple model segments for distributed execution.
These segments are deployed across IoT devices using an over-the-air (OTA) update mechanism for efficient remote distribution. In the IoT layer of Fig.~\ref{fig:arch1}, microcontrollers receive the previously edge-prepared SL models distributed as firmware files. The process starts with the first IoT node that acquires the data from a camera sensor and uses it as input for the CNN model (e.g., MobileNet-V2). It then runs a part of the model for inference. Next, first IoT node assumes the role of a server, while the other nodes in the network act as clients. The intermediate layer outputs are transmitted from one device to the next until a prediction result is generated and sent back to the original IoT node. The exchange of intermediate layer outputs between IoT nodes is performed using lightweight communication protocols, including TCP/UDP over WLAN, as well as low-power alternatives such as ESP-NOW and BLE.

\section{TinyML and Split Learning Solution} \label{sec:solution}

In this section, we focus on the deployment of a TinyML-based CNN model over the previously outlined SL-TinyML framework (Fig.~\ref{fig:arch1}), with the objective of optimal selection of split points to minimize end-to-end inference latency. After the problem formulation, we leverage
the capabilities of the Beam Search-based algorithm to identify the most suitable solutions, which we then use experimentally. 

\subsection{Problem formulation}

Let the model be represented as:
\begin{equation} \label{eq1}
f(x) = f_L \circ f_{L-1} \circ \cdots \circ 
f_1(x)
\end{equation}
where $f_k$ is the function implemented by the $k$-th layer, $x$ is the input, and  $L$ is a total number of layers. Let $S=\{ s=(s_1,..., s_i,...,s_{N-1})| s_i\in [1, N-1], s_i<s_{i+1} \}$, where and $S$ is a set of  split points of a $L$-layered TinyML model between $N$ devices. 
The intermediate activation after split point $s$ at device $i$ : 
\begin{equation}
h_{s_i} = f_{s_i} (h_{s_i-1}),  h_0=x \label{eq:2}
\end{equation}

If the network is split between $N$ devices  at split points $s$, the split model $F$ can be represented as a set of partitions given by: 
\begin{equation}\begin{split}
F=&\{f^1,...,f^i, ..., f^N\},\\
f^i(h_{s_{i-1}}) =& f_{s_{i}} \circ f_{s_{i}-1} \circ \cdots \circ f_{s_{i-1}+1}(h_{s_{i-1}}),\\
& \forall i \in [1, N], \text{ and }s_0=0,  s_N=L
\end{split}
\label{eq:4}
\end{equation}

The round-trip-time latency after the CNN model was split among $N$ IoT devices, assuming different potential split points $s$, consists of device-local processing latency $T_{\text{d}}$ and transmission latency $T_{\text{tr}}$.
Each device-local processing latency $T_{\text{d}_{i}}$ consists of model and input loading $ T_{\text{load}_{i}}$, tensors allocation $T_{\text{ta}_{i}}$, inference $T_{\text{infer}_{i}}$ and intermediate activations buffering $T_{\text{iab}_{i}}$, and is expressed as follows:
\begin{equation}
T_\mathrm{d_{i}}(s) = T_{\text{load}_{i}}(s) + T_{\text{ta}_{i}}(s) + T_{\text{infer}_{i}}(s) + T_{\text{iab}_{i}}(s)
\label{eq:processing}
\end{equation}
with the sum of all device-local processing latencies: 
\begin{equation}
T_\mathrm{d}(s) = \sum_{i=1}^{N} T_{d_{i}}(s)
\label{eq:5}
\end{equation}
The transmission latency $T_{\text{tr}}$ is the sum of all transmission latencies  $T_{\text{tr}_{i,i+1}}$ between devices $i$ and $i+1$ of a split $s$.
\begin{equation}\begin{split}
  T_{\text{tr}}(s, r) =& \sum_{i=1}^{N-1}T_{\text{tr}_{i,i+1}}(s, r)
\end{split}
\label{eq:5}
\end{equation}
For generalization, we consider that $L_{s_i}$ is the size of the output of layer $s_i$, and that packets can be lost with probability $p$. We denote by $T_\text{prop}$ the propagation delay per packet, and $T_\text{ack}$ acknowledgment or protocol overhead per packet. The expected transmission time  $T_{\text{tr}_{i,i+1}}$ is given by:
\begin{equation}\begin{split}
T_{\text{tr}_{i,i+1}}(s, r)=& K_{s_i} \left( \frac{MTU}{r(1-p)} + T_\text{prop} + T_\text{ack} \right)
\end{split}
\label{eq:tr_i}
\end{equation}
where $K_{s_i}=\frac{L_{s_i}}{MTU}$ is the number of packets that can be sent as each protocol has a MTU limit. 


The overall latency can be expressed as follow: 

\begin{equation}
T_\mathrm{inference}(s;r) = T_\mathrm{d}(s) + T_{\text{tr}}(s, r)
\label{eq:inference}
\end{equation}




Based on Eq.\ref{eq:inference}, we formulate the optimization problem as, 
\begin{equation}
s^* = \arg\min_{s \in S} T_\mathrm{inference}(s;r)
\label{eq:opt}
\end{equation}

\subsection{Heuristic Split Point Selection}
\label{subsec:heuristics}

To solve Eq.~\eqref{eq:opt}, we consider three search strategies for split point selection:
\textbf{Beam Search}, \textbf{Greedy Search}, and \textbf{First-Fit Search}.
All three aim to find the split configuration 
$s^* = \{ s_1, \dots, s_{N-1} \}$ that minimizes the expected end-to-end latency
$T_\mathrm{inference}(s;r)$.

An exhaustive evaluation of all possible split combinations requires
$\binom{L-1}{N-1}$ candidates, which becomes computationally infeasible
for even moderate values of $L$. Therefore, we rely on heuristic methods
that provide efficient approximations with significantly reduced complexity.

\subsubsection{Beam Search-Based Split Point Selection}
\begin{algorithm}[h]
\caption{Beam Search for Split Point Optimization}
\label{alg:beamsearch}
\small
\begin{algorithmic}[1]
\Require Number of layers $L$, number of devices $N$, beam width $B$
\Ensure Best split configuration $s^*$
\State Initialize beam $\mathcal{B} \gets \{(0, 0, [])\}$
\For{$k = 1$ to $N$}
    \State $\mathcal{B}_{\text{new}} \gets \emptyset$
    \ForAll{$(pos, cost, splits) \in \mathcal{B}$}
        \For{$next = pos+1$ to $L-(N-k)$}
            \State $c_{\text{seg}} \gets \mathrm{CostSegment}(pos+1, next, k)$
            \State Add $(next, cost + c_{\text{seg}}, splits \cup \{next\})$ to $\mathcal{B}_{\text{new}}$
        \EndFor
    \EndFor
    \State $\mathcal{B} \gets$ top-$B$ elements of $\mathcal{B}_{\text{new}}$ by ascending cost
\EndFor
\State $s^* \gets$ configuration in $\mathcal{B}$ with minimum cost
\State \Return $s^*$
\end{algorithmic}
\end{algorithm}

Beam Search mitigates this by expanding only the top-$B$ most promising partial solutions (beam width $B$) at each step.
At every iteration, it incrementally adds new split points and prunes the search space to retain only a limited number of candidates with the lowest cumulative latency.

At iteration $k$, each candidate represents a partial split configuration
$s_{1:k}$ with cumulative cost:
\begin{equation}
C(s_{1:k}) = \sum_{i=1}^{k} \mathrm{CostSegment}(s_{i-1}+1, s_i, i),
\end{equation}
where $\mathrm{CostSegment}(a,b,i)$ denotes the latency of assigning layers
$[a,b]$ to device $i$, including both local inference and transmission costs.

The algorithm incrementally expands candidates by selecting feasible next split
points $s_k \in [s_{k-1}+1, L-(N-k)]$, and prunes the search space by keeping
only the best $B$ candidates at each step. After $N$ iterations, the configuration
with the minimum total latency is selected.

\subsubsection{Greedy Search}

The Greedy Search algorithm selects split points sequentially by minimizing
the immediate segment cost at each step.
At iteration $k$, given the previous split point $s_{k-1}$, the next split point
is chosen as:
\begin{equation}
s_k = \arg\min_{j \in [s_{k-1}+1,\, L-(N-k)]}
\mathrm{CostSegment}(s_{k-1}+1, j, k).
\end{equation}

This approach has low computational complexity since it avoids exploring
multiple candidates. However, because decisions are based solely on local
optimization, it may not always give the globally optimal solution.

\begin{algorithm}[]
\caption{Greedy Search }
\label{alg:greedysearch}
\small
\begin{algorithmic}[1]
\Require Number of layers $L$, number of devices $N$
\Ensure Split configuration $s$
\State Initialize $pos \gets 0$, $s \gets [\,]$
\For{$k = 1$ to $N-1$}
    \State $best\_next \gets \text{None}$, $best\_cost \gets \infty$
    \For{$next = pos+1$ to $L-(N-k)$}
        \State $c_{\text{seg}} \gets \mathrm{CostSegment}(pos+1, next, k)$
        \If{$c_{\text{seg}} < best\_cost$}
            \State $best\_cost \gets c_{\text{seg}}$
            \State $best\_next \gets next$
        \EndIf
    \EndFor
    \State Append $best\_next$ to $s$
    \State $pos \gets best\_next$
\EndFor
\State \Return $s$
\end{algorithmic}
\end{algorithm}

\subsubsection{First-Fit Search}

The First-Fit Search algorithm selects split points based on a simple acceptance rule,  scanning the feasible split points from left to right and choosing the first one whose segment cost satisfies a predefined latency threshold $\tau_k$ for
device $k$:
\begin{equation}
\mathrm{CostSegment}(s_{k-1}+1, s_k, k) \leq \tau_k.
\end{equation}

If no candidate satisfies the threshold, the algorithm selects the last feasible
split point before violating the device constraint. This method is computationally
efficient and well suited for resource-constrained environments, although its
performance depends on the choice of threshold.

\begin{algorithm}[]
\caption{First-Fit Search }
\label{alg:firstfit}
\small
\begin{algorithmic}[1]
\Require Number of layers $L$, number of devices $N$, thresholds $\tau_k$
\Ensure Split configuration $s$
\State Initialize $pos \gets 0$, $s \gets [\,]$
\For{$k = 1$ to $N-1$}
    \State $chosen \gets \text{False}$
    \For{$next = pos+1$ to $L-(N-k)$}
        \State $c_{\text{seg}} \gets \mathrm{CostSegment}(pos+1, next, k)$
        \If{$c_{\text{seg}} \leq \tau_k$}
            \State Append $next$ to $s$
            \State $pos \gets next$
            \State $chosen \gets \text{True}$
            \State \textbf{break}
        \EndIf
    \EndFor
    \If{not $chosen$}
        \State $fallback \gets L-(N-k)$
        \State Append $fallback$ to $s$
        \State $pos \gets fallback$
    \EndIf
\EndFor
\State \Return $s$
\end{algorithmic}
\end{algorithm}


\section{Experimental Setup and Measurements} 
\label{sec:results}
In this section, we describe the testbed settings, detailing the hardware and software tools used in the experiments. Then, we present measurements and performance evaluation of latency results obtained for wireless communication protocols. Finally, we evaluate the performance of the Beam Search algorithm to determine the best model split points.

\subsection{Testbed Settings}

\begin{figure*}[]
\centering
\resizebox{\textwidth}{!}{%
\includegraphics[scale=0.48]{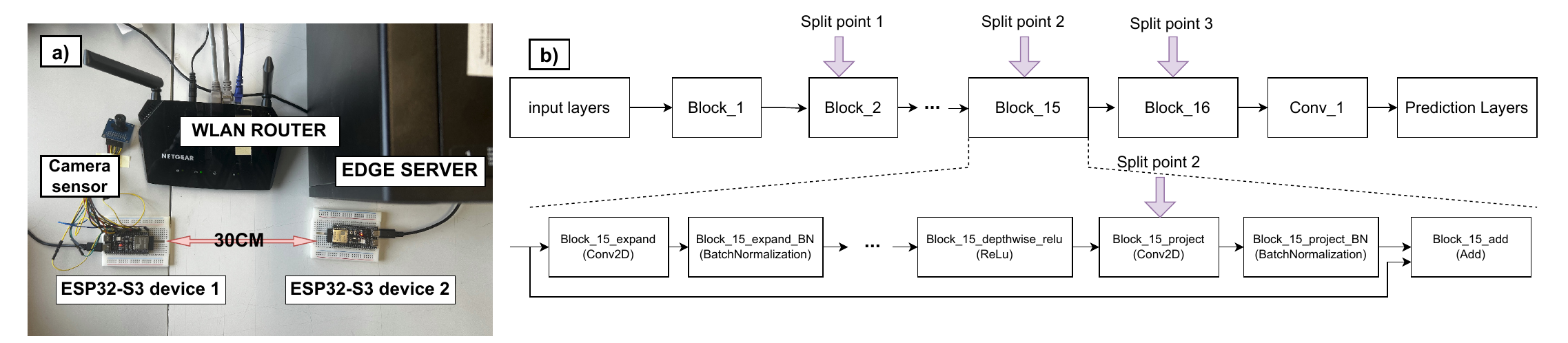}
}
\caption{a) Experimental settings b) Detailed Mobilenet-V2 architecture and split points.}
\label{fig:modelarch}
\end{figure*}

As illustrated in Fig.~\ref{fig:modelarch}.a), the proposed TinyML + SL testbed solution consists of one edge server and two IoT devices. For the edge server, which is used as the development environment for the ML application, we use a Desktop PC (Intel(R) Core(TM) i9-14900 CPU, 64 GB RAM). For IoT devices we use ESP32-S3-WROOM-1 boards (240 MHz, 16 MB Flash, 512 KB SRAM, 8 MB PSRAM, 2.4-GHz Wi-Fi (802.11 b/g/n), Bluetooth 5 LE (2-Mbps
PHY, long-range, mesh) can deploy embedded TinyML models using TFLM.
A camera sensor (OV7670 300KP VGA-Mini).
The full MobileNet-V2 model as lightweight convolutional neural network is loaded with a width multiplier of 0.35. An internal layer (\texttt{block\_16\_project\_BN}) was randomly selected as the split point. This way the model is split into two parts; \emph{Part 1} includes all layers up to the selected split layer, and \emph{Part 2} constructs the remaining layers sequentially. To reduce the model size while minimizing accuracy degradation, we employ quantization using TFLite\footnote{https://www.tensorflow.org/lite}. This ensures alignment of quantization parameters with their respective scale and zero points
\cite{Jacob_2018_CVPR}.
The final result is a pair of quantized models (.tflite) files that were converted into C++ header (.h) and source (.cc) files, suitable for integration into embedded firmware. Two separate Arduino sketches were developed, each embedding one of the model parts compiled into binary firmware files. Our testbed consists of two ESP32-S3 devices, positioned 10 cm apart, connected to an access point to enable access to the OTA update server. This connection allows the devices to upload new firmware versions. The following 
workflow between IoT nodes provides a detailed understanding of the distributed inference process. For both IoT devices, Table~\ref{table:funmodels} summarizes the characteristics of the protocols chosen for the implementation for the corresponding libraries used in our code\footnote{https://github.com/tubs-kns/tiny}, along with the connection types and maximum payload size for the IoT devices.

\begin{table}[ht]
\centering
\resizebox{\columnwidth}{!}{
\begin{tabular}{|l|c|c|c|}
\hline
\textbf{Protocol} & \textbf{Max Devices} & \textbf{Max Payload} & \textbf{Connection Type} \\ \hline
ESP-NOW & 20 peers    & 250 B  & Peer-to-peer (MAC/STA) \\ \hline
UDP     & Unlimited   & 1472 B & Connectionless (AP/STA) \\ \hline
TCP     & 4--10       & 1460 B & Connection-oriented (client/server) \\ \hline
BLE     & 7 (classic) & 512 B  & Connection-oriented (GATT) \\ \hline
\end{tabular}}
\caption{Protocol characteristics for IoT-based split inference.}
\label{table:funmodels}
\end{table}

   



\textit{IoT device 2} continuously listens for incoming activation chunks, reconstructing the full 
tensor before executing the second model partition. The resulting classification probabilities are 
formatted and transmitted back to IoT device 1. Latency is measured via high-resolution ESP32-S3 
timers, with RTT timestamps captured at each operation boundary: protocol setup, model loading, 
tensor allocation, inference, activation buffering, transmission, and feedback reception. Firmware 
is open source and fully reproducible.

\noindent\textbf{Note on experimental scope:} Measurements are conducted on the two-device testbed described above. All results for $N > 2$ devices are obtained from a  model-based simulation, with parameters used from the experimental measurements per-layer inference and transmission costs.

\subsection{Measurement setup and results}


\begin{table*}[ht]
\centering
\footnotesize                       
\resizebox{\textwidth}{!}{   
\begin{tabular}{|c|c|c|c|c|c|c|c|}
\hline
\multicolumn{2}{|c|}{\textbf{Layers} (Output shape)} & 
\multicolumn{2}{c|}{\textbf{block\_2\_expand} (None,56,56,48)} & 
\multicolumn{2}{c|}{\textbf{block\_15\_project} (None,7,7,56)} & 
\multicolumn{2}{c|}{\textbf{block\_16\_project\_BN}(None,7,7,112)} \\ \hline
\multicolumn{2}{|c|}{\textit{Model size (D1, D2) }} & 
\multicolumn{2}{c|}{752.6 kB, 11.8 MB} & 
\multicolumn{2}{c|}{2.2 MB, 9.7 MB} & 
\multicolumn{2}{c|}{2.7 MB, 9.2 MB} \\ \hline
\textit{Protocol} & \textit{Bytes per chunk} & \textit{Latency} & \textit{N. packets} & 
\textit{Latency} & \textit{N. packets} & 
\textit{Latency} & \textit{N. packets}  \\ \hline
\multirow{3}{*}{UDP}
                     & 1472 & 145.1 ms & 103  & 2.26 ms & 2  & 5.2  ms & 4  \\ \cline{2-8}   
                     & 1460 & 83.9 ms & 104  & 1.4 ms & 2 &  3.2 ms & 4 \\ \cline{2-8}
                     & 1200 &  98,3 ms & 126  & 2.2  ms & 3  & 3.7  ms & 5  \\ \hline
\multirow{3}{*}{TCP} 
                     & 1472 & 558.7 ms & 103  & 8.6  ms & 2  & 19.2   ms & 4  \\ \cline{2-8}
                     & 1460 & 563.3 ms & 104 & 8.5  ms & 2  & 19.3   ms & 4  \\ \cline{2-8}
                     & 1200 & 393.9 ms & 126  & 8.8  ms & 3  & 15.719 ms & 5  \\ \hline
ESP-NOW                &  250 & 1897.0 ms & 603 & 34.6 ms & 11  & 69.2  ms & 22 \\ \hline
BLE                  &  512 & 7305.94 ms & 603  & 148.9 ms & 11  & 272.9 ms & 11  \\ \hline
\end{tabular}}
\caption{Benchmarking internode transmission time in (ms) of different model splits of MobileNet-V2}
\label{table:splits}
\end{table*}

\begin{table}[ht]
\centering
\footnotesize 
\resizebox{0.45\textwidth}{!}{
\tiny
\begin{tabular}{|l|c|c|}
\hline

 \multicolumn{1}{|c|}{\textbf{Parameter}}             & \textbf{Device 1}  & \textbf{Device 2} \\ \hline
 \multirow{1}{*}{Model loading }           &  0.0001 ms & 0.01 ms\\  \hline
\multirow{1}{*}{Input loading}                   &   9.8 ms& 0.0001 ms\\ \hline
\multirow{1}{*}{Tensors allocation}                   & 43.00 ms&  10.00 ms  \\ \hline
\multirow{1}{*}{Inference}                   & 3053.75  ms& 437.00 ms \\ \hline
\multirow{1}{*}{Intermediate activations buffering }                   &  0.02 ms & --  \\ \hline
\end{tabular}
}
\caption{Benchmarking the processing time (in ms) of TinyML parameters on the 2 devices (Server-Client)}
\label{table:delays}
\end{table}

\begin{table}[h]

\resizebox{\linewidth}{!}{
\begin{tabular}{|l|c|c|c|c|c|c|}
\hline

 \multicolumn{1}{|c|}{\textbf{Parameter}}             & \textbf{UDP} & \textbf{TCP} & \textbf{ESP-NOW }  & \textbf{BLE}   \\ \hline
 \multirow{1}{*}{Protocol setup}           & 2134.9 ms  &  2590.623 ms & \textbf{48.00 ms}  &  6378.52 ms\\  \hline
\multirow{1}{*}{Feedback delay}                   &   \textbf{0.649 ms }   & 2.645 ms  &  1.115 ms & 24.550 ms \\ \hline
\multirow{1}{*}{RTT}                   &    5800.0 ms   & 6202.2 ms  & \textbf{3662 ms} & 10443.55 ms\\ \hline 
\end{tabular}
}
\caption{Benchmarking the processing time (in ms) of communication protocols parameters between 2 devices (Server-Client)}
\label{table:protocols}
\end{table}

Table~\ref{table:splits} evaluates distributed inference using MobileNet-V2, analyzing the effect of split point and communication protocol on end-to-end latency. The detailed architecture of MobileNet-V2 is illustrated in Figure \ref{fig:modelarch}.b), highlighting its main building blocks and the locations of the model splits. Three split points are randomly chosen at \textit{Block 2 expand}, \textit{Block 15 project}, and \textit{Block 16 project BN}, to analyze the protocol performance trends.


We evaluated the latency and number of packets for four wireless communication methods UDP and TCP over WiFi, ESP-NOW, and BLE.  UDP achieves the lowest transmission latency (1.4 ms, 2 packets)  particularly when used with later model splits (e.g., `\textit{Block 15 project layer}`), owing to its large MTU (1460 bytes). BLE produces the highest latency (7305.94 ms, 603 packets \textit{Block 2 expand layer} due to excessive fragmentation and increased overhead during transmission from its 512-byte MTU.

ESP-NOW outperforms BLE despite a 250-byte 
packet limit, as its MAC-layer broadcast mechanism eliminates connection-handshake overhead. TCP incurs higher latency than UDP due to acknowledgment and retransmission overhead. The anomalous latency pattern observed in \textit{Block 2 expand layer} for TCP is attributed to the large number of packets required to transmit the high-dimensional intermediate activations ($\approx $150 KB). Additionally, ESP32's limited internal TCP buffer and flow control mechanisms cannot efficiently handle continuous transmission of over 100 large packets, leading to stalls and increased round-trip time \cite{s20174774}. In contrast, smaller chunks (e.g., 1200 bytes) reduce retransmission impact and better align with TCP's congestion window, resulting in lower overall latency.
Reducing the \textit{chunk size} (e.g., from 1460 to 1200 bytes in UDP) led to a greater number of packets and marginal increases in transmission time.

Tables~\ref{table:delays} and~\ref{table:protocols} detail processing and communication delays of the proposed framework at \textit{Block 16 project BN}. As expected, the computational workload on IoT device 1 is significantly higher, particularly in inference time (3053.7ms), compared to IoT device 2 (437 ms),which only performs classification tasks. Communication between the two devices for transmitting intermediate activations and receiving feedback is strongly influenced by the choice of wireless protocol. ESP-NOW achieves the best RTT (3.6 s) due to negligible setup time (48 ms) and low 
feedback delay (1.1 ms), despite higher per-payload transmission cost. BLE results in the worst RTT 
(10.4 s) due to its high setup and feedback times, 
while UDP and TCP both incur $>$2 s protocol setup overhead.


\begin{figure*}[h]
    \centering
    \subfloat[\footnotesize Mobilenet-V2]{%
       \includegraphics[scale=0.3]{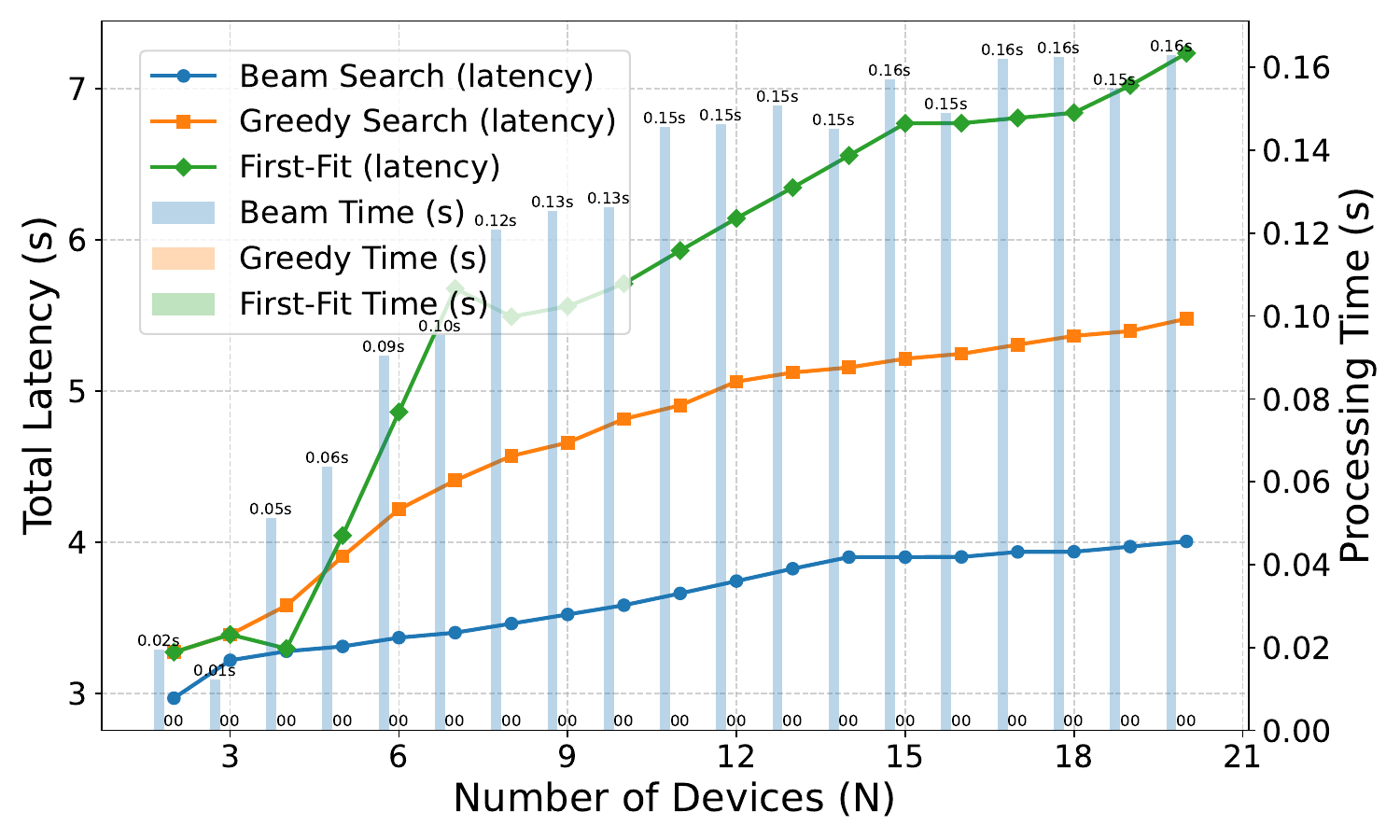}
        \label{fig:middle}%
    }
    \subfloat[\footnotesize ResNet50]{%
       \includegraphics[scale=0.3]{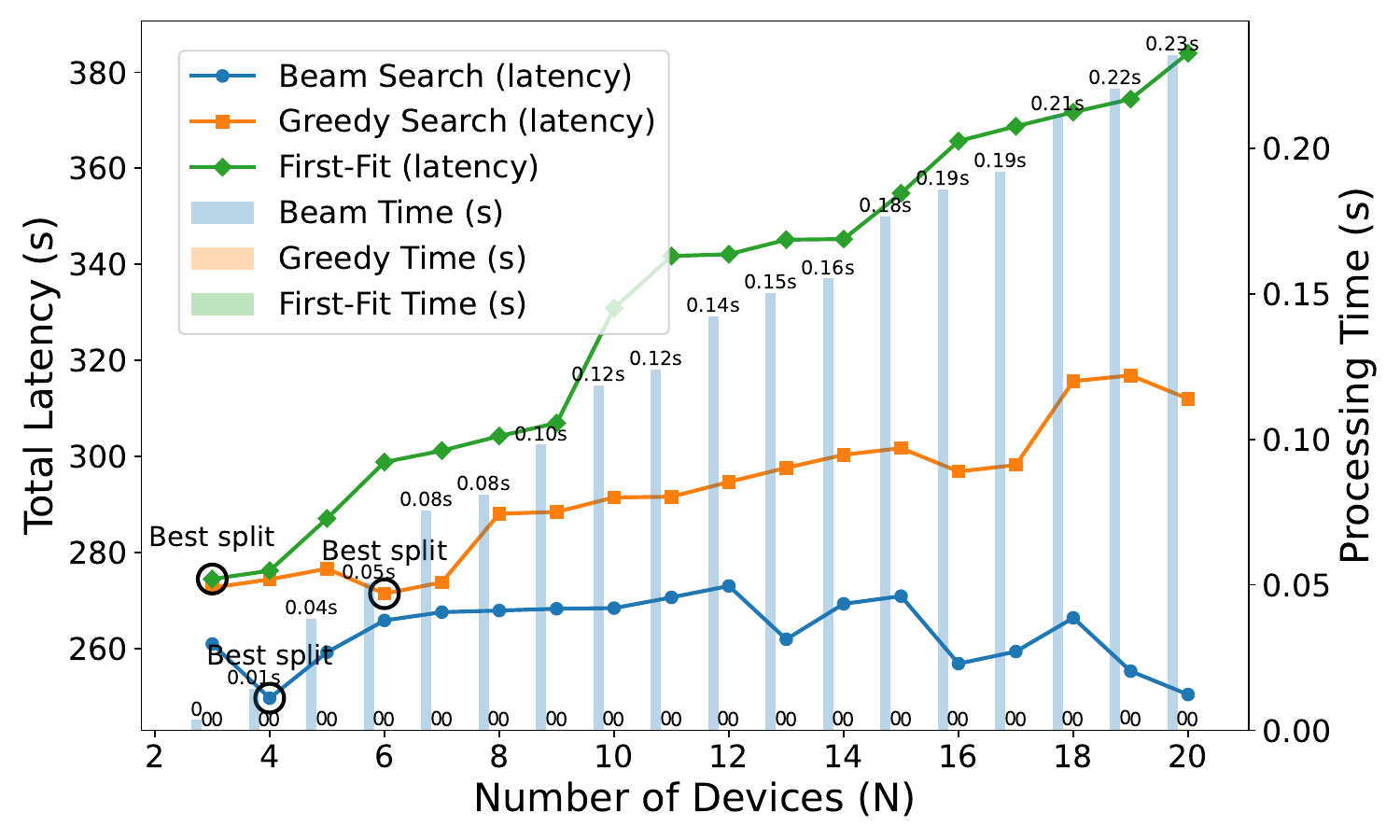}
        \label{fig:right}%
    }
    \caption{Benchmarking latency and processing time vs number of devices for heuristic algorithms. }
    \label{fig:PerformanceTinyMLInf}
\end{figure*}
\subsection{Optimization with Heuristic based Selection}

We benchmark inference and transmission latency at every intermediate layer ($\forall i \in [1,L]$), 
using measured per-layer costs, packet loss $p$, and MTU as inputs to the heuristic algorithms. The algorithms aim to optimize the selection of the best split points for minimizing end-to-end device inference latency with N devices (Results for N > 2 devices are obtained via model-based simulation using experimentally measured per-layer latency and transmission costs).

\begin{figure}[ht]
\centering
\includegraphics[width=0.8\linewidth]
{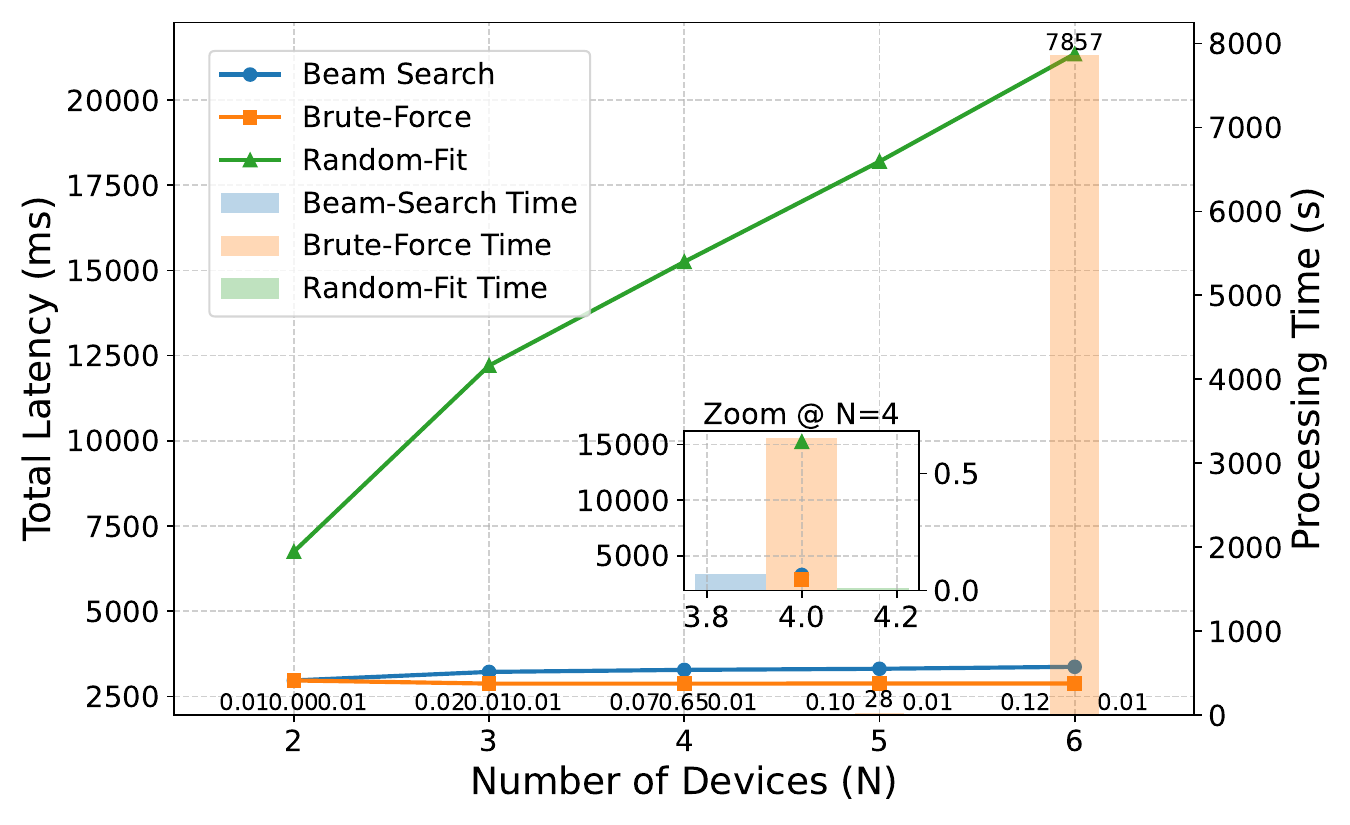}
\caption{Benchmarking latency and processing time vs number of devices.}
\label{fig:BS}
\end{figure}

Figure~\ref{fig:BS} compares Beam Search with Brute-Force and Random-Fit, illustrating the resulting end-to-end latency as a function of the number of devices, along with the processing time required to execute each algorithm, using data from the experimental benchmarking.
Brute-Force provides  marginally lower latency but its processing time grows exponentially, reaching $\approx$7857 s for 6 devices, because it explores every possible configuration exhaustively. Beam Search achieves comparable latency in $\approx$0.06 s, and reduces latency by $>$600\% over Random-Fit for 6 devices, making it the practical choice for real-time deployments.

Next, we compared the three proposed heuristics, as shown in Figure \ref{fig:PerformanceTinyMLInf}, and evaluated them for two ML models of small and large sizes: Mobilenet-V2 and ResNet50, used for image classification. The results show that Beam Search consistently achieves the lowest latency across both models and all device counts.  
Greedy Search exhibits higher latency due to its locally optimal decisions.  First-Fit provides the highest latency, as its simplified selection strategy leads to less efficient split placement.
ResNet50 shows latency fluctuation at higher device counts due to incapacity of nodes to run some model segments, unlike MobileNet-V2 where all split points remain valid. In terms of processing time, all methods remain highly efficient and scalable, staying below 0.17 s for MobileNet-V2 and 0.23 s for ResNet50, even as the number of devices increases. Beam Search requires slightly more computation than Greedy and First-Fit, but with minimal difference.

 \section{Conclusion}\label{sec:conclusion}
 We experimentally demonstrated for the first time the distributed TinyML inference by pipelining MobileNet-V2 across two ESP32-S3 nodes. We analyzed how split points affect end-to-end latency. ESP-NOW delivered the lowest RTT thanks to minimal setup and MAC-level exchange, but its small packets penalize larger transfers. As expected, UDP provided the lowest transmission latency, while TCP remained reliable yet costly due to setup overhead and buffer driven retransmissions. We conclude that ESP-NOW is the best protocol for end-to-end RTT, owing to its negligible setup overhead, despite UDP achieving lower raw transmission latency for larger payloads. We proposed a mathematical formulation and Beam Search based algorithm to find the best model split point to minimize the end-to-end latency. The Beam Search algorithm is scalable for large number of devices, unlike brute-force method, providing practical solution for TFLite capable hardware. Future work will build a dynamic, adaptive framework that selects protocols, activation chunk sizes, and split points at runtime based on network conditions, and device resources.

\vspace{-0.1cm}
\bibliographystyle{IEEEtran}
\bibliography{mybib}

\end{document}